\documentclass[journal]{IEEEtran}
\usepackage{textcomp}
\usepackage[noadjust]{cite}
\usepackage[T1]{fontenc} 
\usepackage{amsmath}
\usepackage[cmintegrals]{newtxmath}
\usepackage{bm}
\usepackage{array}
\usepackage{url}
\usepackage{amsmath}
\usepackage{booktabs}
\usepackage{multirow}
\usepackage{graphicx}
\usepackage{subcaption}
\usepackage{balance}

\ifCLASSINFOpdf
\else
\fi

\begin{document}

\title{A Nonlinear Spectral Approach for Radar-Based Heartbeat Estimation via Autocorrelation \\of Higher Harmonics}

\author{Kohei~Shimomura,
        Chi-Hsuan~Lee,
        and~Takuya~Sakamoto,~\IEEEmembership{Senior Member,~IEEE}
\thanks{K.~Shimomura and T.~Sakamoto are with the Department of Electrical Engineering, Graduate School of Engineering, Kyoto University, Kyoto 615-8510, Japan.}
\thanks{C.~-H.~Lee is with the Quanta Computer Inc., Taoyuan City 33377, Taiwan.}}


\maketitle

\begin{abstract}
This study presents a nonlinear signal processing method for accurate radar-based heartbeat interval estimation by exploiting the periodicity of higher-order harmonics inherent in heartbeat signals. Unlike conventional approaches that employ selective frequency filtering or track individual harmonics, the proposed method enhances the global periodic structure of the spectrum via nonlinear correlation processing. Specifically, smoothing and second-derivative operations are first applied to the radar displacement signal to suppress noise and accentuate higher-order heartbeat harmonics. Rather than isolating specific frequency components, we compute localized autocorrelations of the Fourier spectrum around the harmonic frequencies. The incoherent summation of these autocorrelations yields a pseudo-spectrum in which the fundamental heartbeat periodicity is distinctly emphasized. This nonlinear approach mitigates the effects of respiratory harmonics and noise, enabling robust interbeat interval estimation. Experiments with radar measurements from five participants demonstrate that the proposed method reduces root-mean-square error by $20$\% and improves the correlation coefficient by $0.20$ relative to conventional techniques.
\end{abstract}

\begin{IEEEkeywords}
radar, nonlinear, harmonics, heart rate, heartbeat, spectrum, autocorrelation, millimeter wave
\end{IEEEkeywords}

\IEEEpeerreviewmaketitle

\section{Introduction}

\IEEEPARstart{A}{ccurate} heartbeat monitoring is critical for health assessment and the detection of potential abnormalities~\cite{10.1109/TIM.2017.2669699}. Although electrocardiography provides precise measurements, it requires direct skin contact, limiting its suitability for continuous monitoring in daily life~\cite{10.1109/ACCESS.2019.2921240, 10.1109/JSEN.2023.3250500}. Yet, non-contact methods such as camera-based systems raise privacy concerns~\cite{10.1109/JBHI.2020.3018394, 10.1109/JBHI.2023.3345486}. In this context, radar-based physiological monitoring has emerged as a promising solution for non-invasive, privacy-preserving heartbeat measurement.

Radar-based heartbeat sensing exploits the periodic modulation of radar echoes generated by minute body movements. However, heartbeat-induced motion is substantially smaller than respiration-induced motion~\cite{10.1109/TIM.2023.3267348}, reducing estimation accuracy. In addressing this issue, various methods have been proposed to enhance heartbeat components through differential operations and mode decomposition~\cite{10.1109/JERM.2018.2879452,10.1109/JSEN.2024.3351274,10.1109/TIM.2020.2978347,10.1109/ACCESS.2025.3575932}. Alternatively, methods leveraging heartbeat harmonics have been shown to improve estimation robustness~\cite{10.1109/EMBC.2014.6944065, 10.1109/ACCESS.2020.2976104, 10.1109/TIM.2025.3555689,10.1109/ICCECE51280.2021.9342280, 10.1109/TMTT.2024.3430513}, as these components are less affected by respiratory interference. However, conventional techniques typically focus on fundamental and lower-order harmonics and do not fully exploit higher-order components, leaving them vulnerable to respiratory artifacts and limiting overall accuracy and robustness.

To overcome these limitations, we propose a nonlinear approach that enhances the global periodicity of the heartbeat spectrum by leveraging higher-order harmonics. After applying smoothing and second-derivative operations to suppress noise and emphasize harmonics, we compute a localized autocorrelation on the spectral regions surrounding heartbeat harmonics. Incoherent summation of these correlations yields a pseudo-spectrum termed the nonlinear harmonic spectrum (NLHS) that accentuates the fundamental periodicity of the heartbeat. This method fundamentally differs from conventional approaches by exploiting the nonlinear characteristics of periodicity enhancement, enabling robust and accurate heartbeat interval estimation without reliance on machine learning or adaptive modeling. Experiments involving radar measurements for five participants demonstrate that the proposed method achieves higher accuracy than existing techniques.

\section{Radar Signal Processing}
\subsection{Estimation of Body Displacement}
A frequency-modulated continuous wave (FMCW) radar system equipped with a linear antenna array is used in this study. Following FMCW demodulation and array signal processing, the complex radar image $I'_\mathrm{C}(t,\bm{r})$ is obtained, where $t$ denotes slow time and $\bm{r}$ is the position vector. To suppress static clutter, the time-averaged component of $I'_\mathrm{C}(t,\bm{r})$ is subtracted, yielding the clutter-suppressed complex radar image $I_\mathrm{C}(t,\bm{r})=I'_\mathrm{C}(t,\bm{r})-(1/T)\int_{0}^{T}I'_\mathrm{C}(t,\bm{r}) \mathrm{d}t$, where $T$ is the total measurement time. The power radar image is computed as $I_\mathrm{P}(t,\bm{r})=\left|I_\mathrm{C}(t,\bm{r}) \right|^2$, and its time average is given by $I_\mathrm{A}(\bm{r})=(1/T)\int_{0}^{T}I_\mathrm{P}(t,\bm{r}) \mathrm{d}t$.
The target position $\bm{r_0}$ is identified from $I_\mathrm{A}(\bm{r})$, and the corresponding complex radar signal is extracted as $s_\mathrm{IQ}(t) = I_\mathrm{C}(t,\bm{r_0})$. Finally, body displacement $d(t)$ is estimated from the phase of $s_\mathrm{IQ}(t)$ as $d(t)=(\lambda/4\pi)\mathrm{unwrap}[\angle s_\mathrm{IQ}(t)]$, where $\lambda$ is the radar wavelength, $\mathrm{unwrap[\cdot]}$ denotes the phase unwrapping operation, and $\angle$ denotes the phase of a complex number.

\subsection{Proposed Method}
To improve estimation accuracy under respiratory interference, the proposed method focuses on using heartbeat harmonics. An NLHS is generated on the basis of a local autocorrelation function computed using only the frequency components near the heartbeat harmonics. First, the displacement signal $d(t)$ is smoothed via convolution with a Gaussian function with a width of 0.1 s. Subsequently, the second derivative $d''(t)$ is computed to enhance heartbeat harmonics. To mitigate high-frequency noise amplification, an improved differentiator based on least-squares smoothing~\cite{differentiator} is employed. This differentiator is expressed as $d''(t) \simeq \{(d_3+d_{-3})+2(d_2+d_{-2})-(d_1+d_{-1})-4d_0\}/{16t_0^2}$, where $t_0$ is the sampling interval, and $d_i=d(t+it_0)$ is defined for $i=-3, -2, \cdots, 3$.

\begin{figure}[tb]
  \centering
  \begin{minipage}[b]{0.48\linewidth}
    \centering
    \includegraphics[width=\linewidth,pagebox=cropbox,clip]{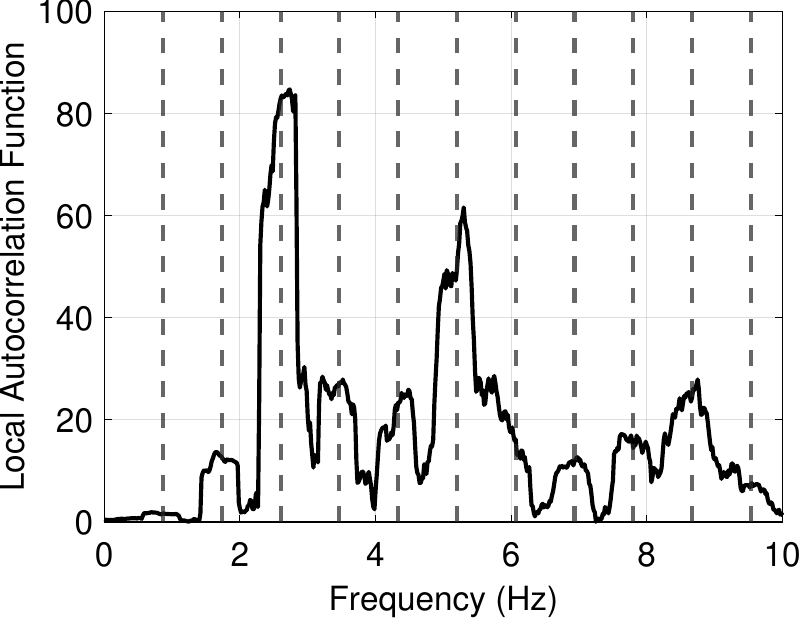}
    \subcaption{}
  \end{minipage}
  \hfill
  \begin{minipage}[b]{0.48\linewidth}
    \centering
    \includegraphics[width=\linewidth,pagebox=cropbox,clip]{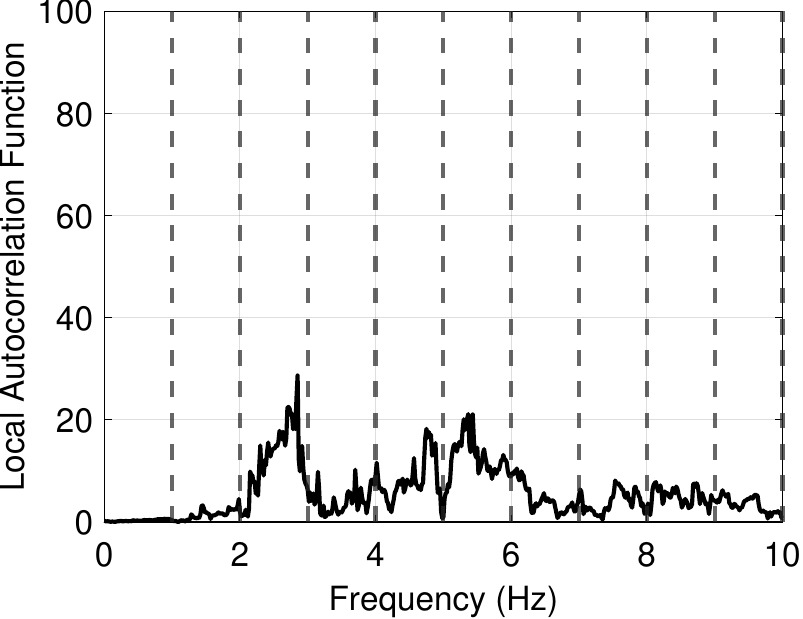}
    \subcaption{}
  \end{minipage}
  \caption{Examples of $c_i(f_0,\Delta f)$:  (a) when $\Delta f$ matches the heartbeat frequency and (b) when $\Delta f$ does not match the heartbeat frequency.}
  \label{fig:ICF}
\end{figure}

To extract periodic structures, the norm of a local autocorrelation function of the Fourier transform $D_i(f)=\mathcal{F}[d_i(t)]$ is calculated as 
\begin{align}
    c(f_0, \Delta f)=\left| \int_{-F/2}^{F/2} D(f_0 + \Delta f + f') D^*(f_0 + f') \mathrm{d}f'\right|,
\end{align}
where $F=0.5$ Hz is the frequency range used for the calculation, and $^*$ denotes the complex conjugate. Fig.~\ref{fig:ICF} (a) and (b) shows examples of $c(f_0, \Delta f)$ when $\Delta f$ matches or does not match the actual heartbeat frequency ($0.87$~Hz), with dashed lines indicating $n\Delta f$ for $n=1,2,\cdots$.

These results indicate that when $\Delta f$ matches the heartbeat frequency, $c(f_0, \Delta f)$ exhibits high values at $f_0=n\Delta f$. In contrast, when $\Delta f$ does not match the frequency of the heartbeat, $c(f_0, \Delta f)$ remains small even at $f_0=n\Delta f$. Using this property, the proposed NLHS is defined as 
\begin{align}
    C_\mathrm{NLHS}(f) = \sum_{n=1}^N c(nf,f),
\end{align}
where $N$ denotes the maximum harmonic order in the analysis.

\begin{figure}[tb]
  \centering
  \begin{minipage}[b]{0.48\linewidth}
    \centering
    \includegraphics[width=\linewidth,pagebox=cropbox,clip]{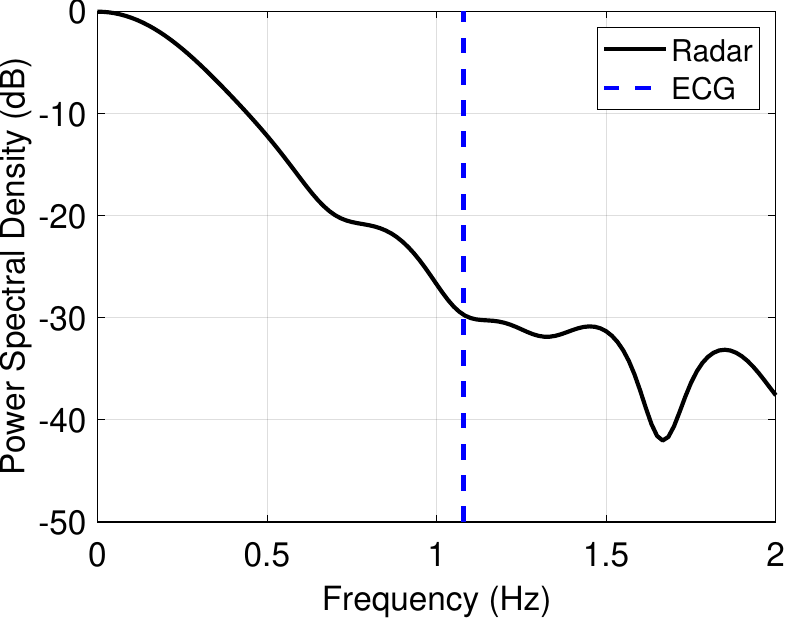}
    \subcaption{}
  \end{minipage}
  \hfill
  \begin{minipage}[b]{0.48\linewidth}
    \centering
    \includegraphics[width=\linewidth,pagebox=cropbox,clip]{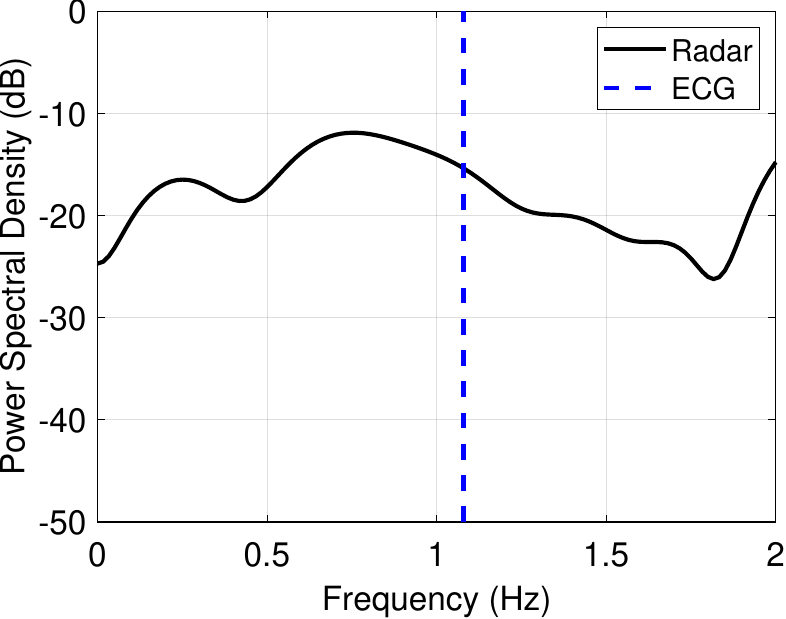}
    \subcaption{}
  \end{minipage}
  \begin{minipage}[b]{0.48\linewidth}
    \centering
    \includegraphics[width=\linewidth,pagebox=cropbox,clip]{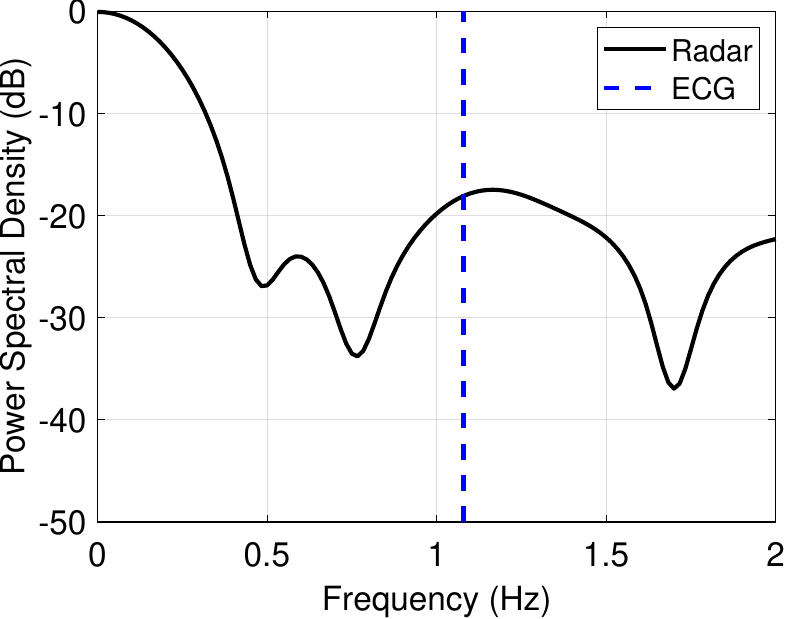}
    \subcaption{}
  \end{minipage}
  \hfill
  \begin{minipage}[b]{0.48\linewidth}
    \centering
    \includegraphics[width=\linewidth,pagebox=cropbox,clip]{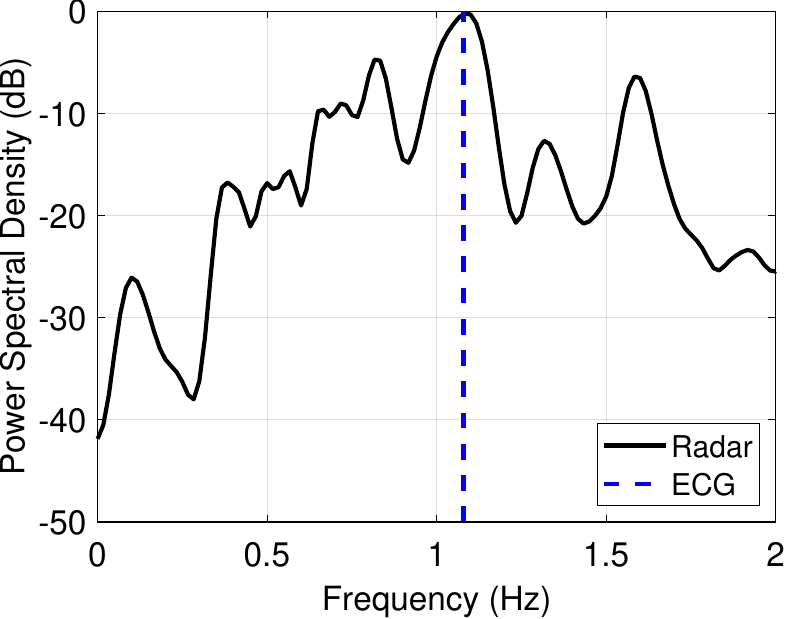}
    \subcaption{}
  \end{minipage}

  \caption{Examples of the power spectral density of (a) $\left| \mathcal{F}[d(t)] \right|^2$, (b) $\left| \mathcal{F}[(\mathrm{d}^2/\mathrm{d}t^2)d(t)] \right|^2$, and (c) $\left| \mathcal{F}[(\mathrm{d}^2/\mathrm{d}t^2)s_\mathrm{IQ}(t)] \right|^2$, and (d) $C_\mathrm{NLHS}(f)$.}
  \label{fig:spectra}
\end{figure}

Fig.~\ref{fig:spectra} (a), (b), and (c) shows the power spectral density functions of $d(t)$, $(\mathrm{d}^2/\mathrm{d}t^2)d(t)$, and $\left|(\mathrm{d}^2/\mathrm{d}t^2)s_\mathrm{IQ}(t)\right|$, respectively.
Fig.~\ref{fig:spectra} (d) presents the proposed NLHS, $C_\mathrm{NLHS}(f)$. In Fig.~\ref{fig:spectra} (a), there is no distinct peak corresponding to the heartbeat. Although differentiation emphasizes higher-frequency components, Fig.~\ref{fig:spectra} (b) still lacks a clear heartbeat-related peak. In Fig.~\ref{fig:spectra} (c), there is a broad and non-sharp peak corresponding to the heartbeat frequency, indicating that the absolute value of the second derivative of the complex radar signal $s_\mathrm{IQ}(t)$ enhances the heartbeat component while avoiding phase unwrapping errors. However, the broadness of this peak suggests limited accuracy. In contrast, the proposed NLHS $C_\mathrm{NLHS}(f)$ in Fig.~\ref{fig:spectra} (d) exhibits a sharper and more distinct peak than the conventional spectra in Fig.~\ref{fig:spectra} (a)--(c), demonstrating the superior performance of the proposed method.

The heartbeat frequency is estimated from the proposed NLHS $C_\mathrm{NLHS}(f)$ as
\begin{align}
    \hat{f}=\underset{f_\mathrm{min}\leq f \leq f_\mathrm{max}}{\arg\max}C_\mathrm{NLHS}(f),
\end{align}
where $f_\mathrm{min} = 0.8$~Hz and $f_\mathrm{max} = 1.7$~Hz respectively represent the typical lower and upper bounds of the human heart rate. The estimated heartbeat interval $\hat{t}'$ is then obtained as the reciprocal of $\hat{f}$.

Because the accuracy of $\hat{t}'$ depends on the choice of $N$, the NLHS is calculated for $N = N_1$ and $N = N_2$, defined as
\begin{align}
    N_1 &= \underset{N_\mathrm{min}\leq N \leq N_\mathrm{max}}{\arg\min}v_N, \\
    N_2 &= \underset{N_\mathrm{min}\leq N \leq N_\mathrm{max}, N\neq N_1}{\arg\min}v_N,
\end{align}
where the variance $v_N = \langle (\hat{t}' - \langle \hat{t}' \rangle)^2 \rangle$ is calculated using the expectation operator $\langle \cdot \rangle$. The empirical bounds are set as $N_\mathrm{min} = 6$ and $N_\mathrm{max} = 15$. As these equations indicate, $N$ is selected to minimize the variance $v_N$ of the estimated intervals. The estimates for $N = N_1$ and $N = N_2$ are denoted as $\hat{t}'_{N_1}$ and $\hat{t}'_{N_2}$, respectively.

Subsequently, inaccurate estimates are excluded according to the criterion
\begin{align}
    \hat{t}=
    \begin{cases}
        \hat{t}'_{N_1} &|\hat{t}'_{N_1} - \hat{t}'_{N_2}| \leq t_\mathrm{\theta}, \\
        \text{NaN} & \text{otherwise},
    \end{cases}
\end{align}
where $t_{\theta} = 10$~ms is a threshold parameter. Finally, a Hampel filter is applied to remove outliers.

Table~\ref{tab:method} summarizes the definitions of the conventional and proposed methods. Conv1A, Conv1B, and Prop1 use the second derivative of the body displacement $d(t)$ as the input signal, whereas Conv2A, Conv2B, and Prop2 employ the absolute value of the second derivative of the complex radar signal $s_\mathrm{IQ}(t)$, following conventional practice~\cite{10.1109/ACCESS.2025.3575932}, which is known to produce clear peaks at the fundamental heartbeat frequency. For Conv1A and Conv2A, the short-time Fourier transform (STFT) is applied. Conv1B and Conv2B adopt the SSA-VHPS method~\cite{10.1109/TMTT.2024.3430513}, which combines specific spectrum analysis (SSA) and a variable harmonic product spectrum (VHPS) to mitigate respiratory harmonic interference in heartbeat estimation.

\begin{table}[tb]
  \centering
  \caption{Definitions of heartbeat estimation methods}
  \begin{tabular}{ccc}
    \toprule
    Method & Input & Estimation \\
    \midrule
    Conv1A& \multirow{3}{*}{$d(t)$} & STFT \\
    Conv1B &  & SSA-VHPS \\
    Prop1  &  & NLHS \\
    \midrule
    Conv2A & \multirow{3}{*}{$\left| \frac{\mathrm{d}^2}{\mathrm{d}t^2} s_\mathrm{IQ}(t) \right| $} & STFT \\
    Conv2B &  & SSA-VHPS \\
    Prop2  &  & NLHS\\
    \bottomrule
  \end{tabular}
  \label{tab:method}
\end{table}

\section{Evaluation of the Estimation Accuracy of the Proposed Method}
\begin{figure}[tb]
  \centering
  \begin{minipage}[b]{0.44\linewidth}
    \centering
    \includegraphics[width=\linewidth,pagebox=cropbox,clip]{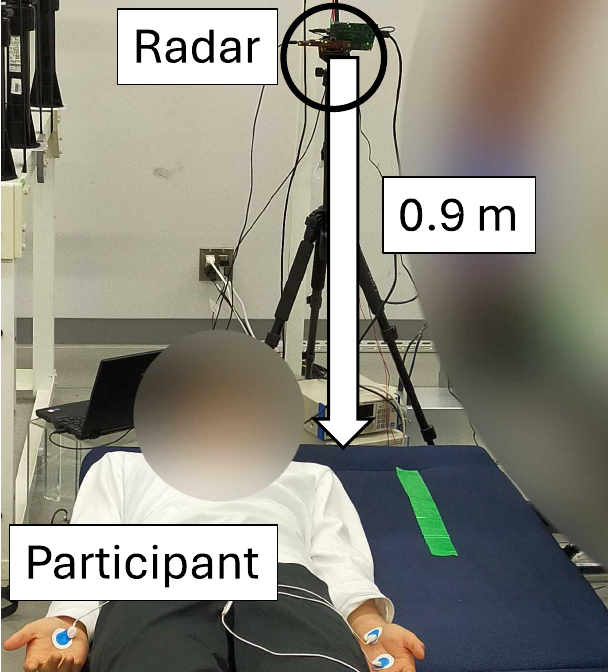}
  \end{minipage}
  \hfill
  \begin{minipage}[b]{0.52\linewidth}
    \centering
    \includegraphics[width=\linewidth,pagebox=cropbox,clip]{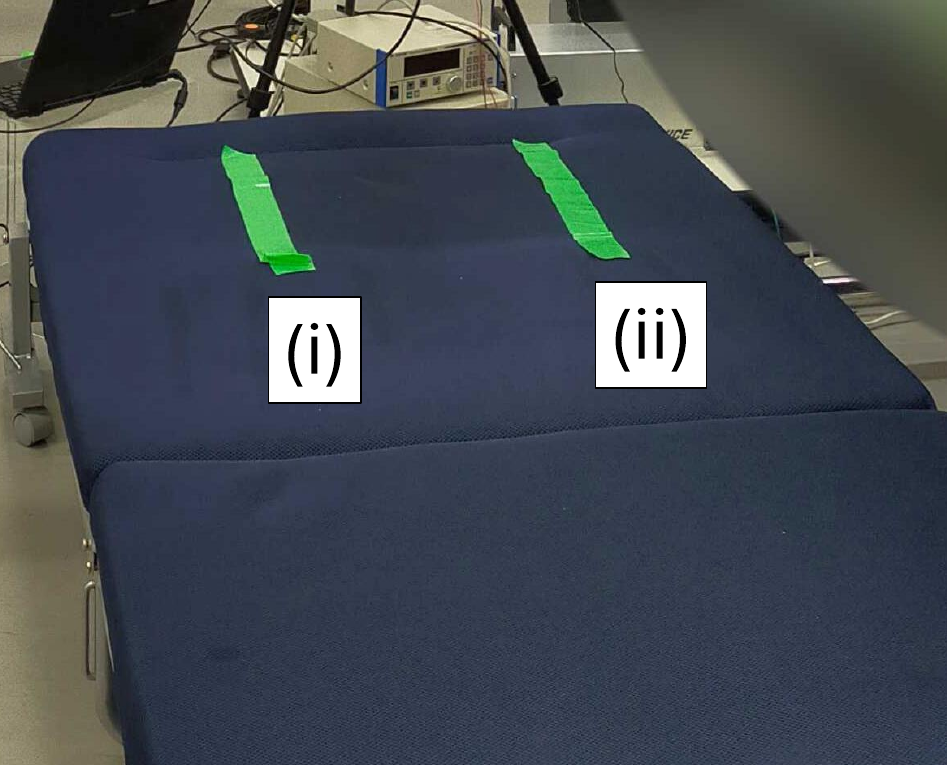}
  \end{minipage}
  \caption{Experimental setup and participant positions.}
  \label{fig:setup}
\end{figure}

An FMCW radar system (IWR6843AOP, Texas Instruments, Dallas, TX, USA) operating in the 60--64~GHz band was employed. This system features a 2D array configuration with three transmitting and four receiving elements. The radar configuration parameters are summarized in Table\ref{tab:radar}. The experimental setup is illustrated in Fig.~\ref{fig:setup}. During the measurements, participants lay in a supine position on a bed while breathing quietly. The radar system was positioned 0.9~m directly above the participant's head. Five participants (four males, one female; aged 22--26 years) took part in the experiment. For each participant, 60-second measurements were conducted twice at two positions (i) and (ii), as shown in Fig.~\ref{fig:setup}. In total, $5 \times 2 \times 2 = 20$ datasets, each with a duration of 60~s, were collected for performance evaluation. To assess the accuracy of the estimated heartbeat intervals, each participant was equipped with an ECG sensor during the measurements, serving as the ground truth reference.

\begin{table}[tb]
  \centering
  \caption{Radar parameters}
  \begin{tabular}{c|c}
    \toprule
    Parameter & Value (unit) \\\midrule
    Start frequency & 60 (Hz) \\
    Frequency slope & 36.21 (MHz/us)\\
    Bandwidth & 3273 (MHz) \\
    Ramp end time & 90.40 (us)\\
    Number of chirps & 4 \\
    ADC sample   & 250 \\ 
    Sampling intervals & 10 (ms) \\
    \bottomrule
  \end{tabular}
  \label{tab:radar}
\end{table}
In this study, the performances of the conventional and proposed methods were evaluated using the root-mean-square error (RMSE), correlation coefficient (CC), and time coverage rate (TCR)~\cite{10.1587/transcom.2020EBP3078}. These metrics are used to assess the agreement between the estimated and actual heartbeat intervals. The RMSE is defined as $\sqrt{\langle |\hat{t} - t_0|^2 \rangle}$, where $t_0$ is the ground-truth heartbeat interval. The CC is given by $\alpha \langle (\hat{t} - \langle \hat{t} \rangle)(t_0 - \langle t_0 \rangle) \rangle$, where $\alpha$ is a normalization factor ensuring CC lies between $-1$ and $1$. The TCR is defined as $m / M$, where $m$ is the number of data segments in which the absolute error of the estimated heartbeat interval is less than a threshold $t_\theta = 30$ms, and $M$ is the total number of segments. Each data segment is 1.0s in duration, and with a total measurement time of 60.0~s per dataset, $M = 60$ in this evaluation.

Table~\ref{tab:rmse} summarizes the RMSE, CC, and TCR averaged over the 20 datasets. Relative to Conv2B, which exhibited the highest accuracy among the conventional methods, the proposed method (Prop1) reduced RMSE by 20\%, increased the CC by 0.20, and improved the TCR by 7.83 points. These results clearly demonstrate the effectiveness of the proposed approach. Furthermore, using $(\mathrm{d}^2/\mathrm{d}t^2)d(t)$ as the input signal yielded superior performance relative to $\left|(\mathrm{d}^2/\mathrm{d}t^2)s_\mathrm{IQ}(t) \right|$. This is because, although $\left| (\mathrm{d}^2/\mathrm{d}t^2)s_\mathrm{IQ}(t) \right|$ enhances the fundamental and low-order heartbeat harmonics, it is less effective at emphasizing higher-order harmonics. In contrast, $(\mathrm{d}^2/\mathrm{d}t^2)d(t)$ more effectively emphasizes higher-order harmonics, indicating that the proposed method benefits from leveraging these components.

\begin{figure}[tb]
  \centering
  \begin{minipage}[b]{0.9\linewidth}
    \centering
    \includegraphics[width=\linewidth,pagebox=cropbox,clip]{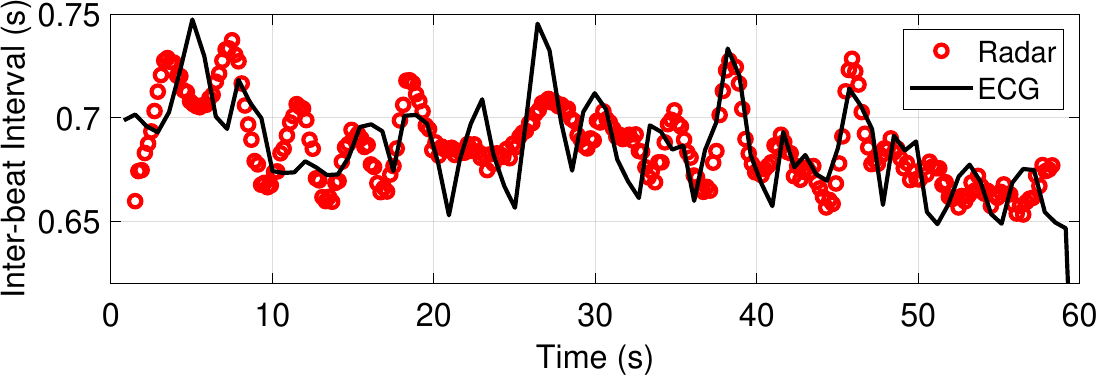}
    \subcaption{}
  \end{minipage}
  \hfill
  \begin{minipage}[b]{0.9\linewidth}
    \centering
    \includegraphics[width=\linewidth,pagebox=cropbox,clip]{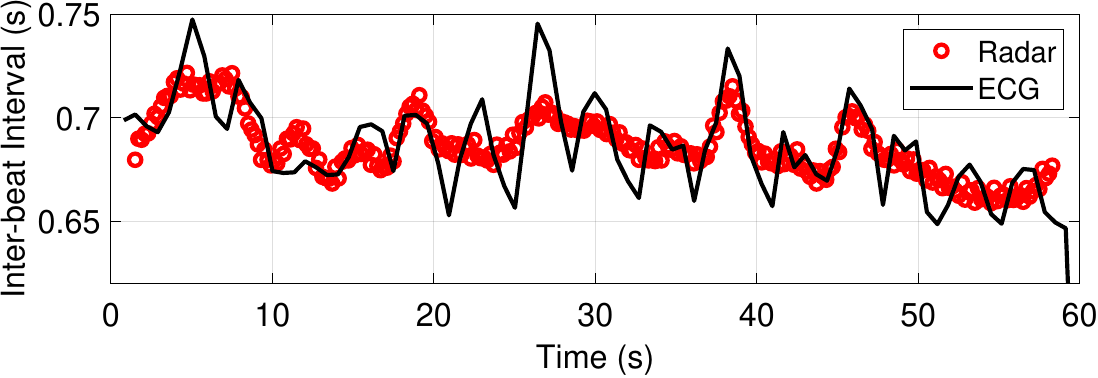}
    \subcaption{}
  \end{minipage}
  \hfill
  \begin{minipage}[b]{0.9\linewidth}
    \centering
    \includegraphics[width=\linewidth,pagebox=cropbox,clip]{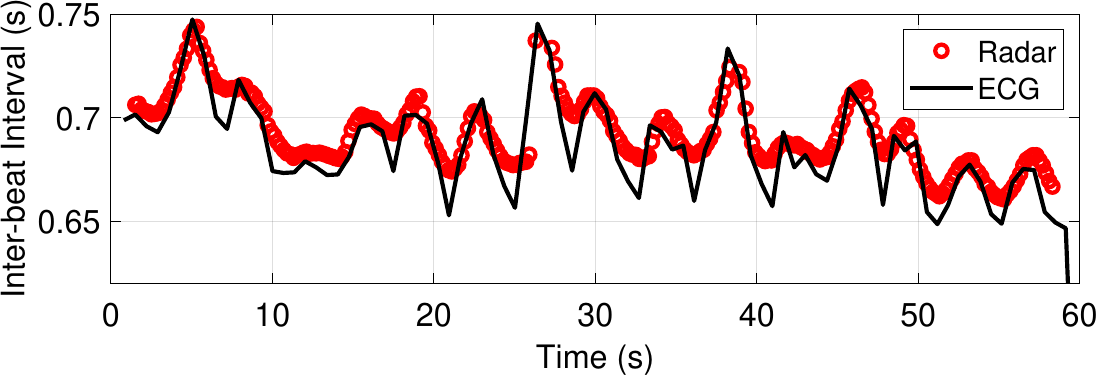}
    \subcaption{}
  \end{minipage}
  \caption{Heartbat interval estimated using (a) Conv2A, (b) Conv2B, and (c) Prop1.}
  \label{fig:IBI}
\end{figure}

Fig.~\ref{fig:IBI} presents a representative example of interbeat interval estimation using both conventional and proposed methods. With Conv2A, the RMSE, CC, and TCR were 16.43ms, 0.65, and 90.00\%, respectively. Conv2B achieved improved values of 13.18ms, 0.77, and 95.00\%. In contrast, Prop1 achieved improvements with an RMSE of 9.46ms, CC of 0.94, and TCR of 96.67\%. These results confirm that the proposed method enables highly accurate heartbeat interval estimation throughout the dataset. This performance improvement is attributed to the use of higher-order heartbeat harmonic components, which are less susceptible to respiratory interference.

\begin{table}
  \centering
  \caption{Heartbeat interval estimation performance}
  \begin{tabular}{c|ccc}
    \toprule
      Method   & RMSE (ms) & CC   & TCR (\%) \\ \midrule
      Conv1A & 156.35    & 0.20 & 27.75 \\
      Conv2A & 60.37     & 0.43 & 48.67 \\
      Conv1B & 216.09    & 0.16 & 25.25 \\
      Conv2B & 52.12     & 0.50 & 56.50 \\
      Prop1  & 41.70     & 0.70 & 64.33 \\
      Prop2  & 82.04     & 0.50 & 47.58 \\
    \bottomrule
  \end{tabular}
  \label{tab:rmse}
\end{table}

\section{Conclusion}
This study proposed a radar-based heartbeat interval estimation method that leverages the harmonic components of heartbeats. The method constructs an NLHS from local autocorrelations of the Fourier transform, focusing on frequency components surrounding the heartbeat harmonics. The heartbeat interval is estimated from the peak of the NLHS. This approach effectively suppresses interference from respiratory harmonics and high-frequency noise. The effectiveness of the proposed method was validated through radar measurements involving five participants. Experimental results demonstrated improved accuracy across all evaluation metrics---the RMSE, CC, and TCR---relative to conventional methods. Additionally, it was observed that posture may affect the estimation accuracy. Future work will address this limitation by collecting data for a broader range of participants and postural conditions.

\section*{Ethics Declarations}
The experimental protocol involving human participants was approved by the Ethics Committee of the Graduate School of Engineering, Kyoto University (permit no. 202223). Informed consent was obtained from all human participants in the study.

\section*{Acknowledgment}
\addcontentsline{toc}{section}{Acknowledgment}
\scriptsize
This work was supported in part by the SECOM Science and Technology Foundation; in part by the Japan Science and Technology Agency under Grant JPMJMI22J2 and Grant JPMJMS2296; in part by the Japan Society for the Promotion of Science KAKENHI under Grant 21H03427, Grant 23H01420, and Grant 23K26115; and in part by the New Energy and Industrial Technology Development Organization. We thank Glenn Pennycook, MSc, from Edanz (https://jp.edanz.com/ac) for editing a draft of this manuscript.
\normalsize


\end{document}